\documentclass[prl,amsmath,amssymb,twocolumn,showpacs,superscriptaddress]{revtex4}
\usepackage{graphicx}

\usepackage{color}
\usepackage{amsmath}
\usepackage{bm}
\newcommand{\eref}[1]{Eq.~(\ref{#1})}

\newcommand{\m}{\mathrm}

\begin{document}

\title{Enhancing optomechanical coupling via the Josephson effect}

\author{T.T.~Heikkil\"a}

\affiliation{Nanoscience Center and Department of Physics, University of Jyv\"askyl\"a, P.O. Box 35 (YFL)
FI-40014 University of Jyv\"askyl\"a, Finland}

\affiliation{Low Temperature Laboratory, Aalto University, P.O. Box 15100, FI-00076 AALTO, Finland}

\author{F. Massel}

\affiliation{Low Temperature Laboratory, Aalto University, P.O. Box 15100, FI-00076 AALTO, Finland}

\affiliation{Department of Mathematics and Statistics, University of Helsinki, FI-00014 Helsinki, Finland}

\author{J. Tuorila}

\affiliation{Department of Physics, University of Oulu, P.O. Box 3000, FI-90014
University of Oulu, Finland}

\author{R. Khan}

\affiliation{Low Temperature Laboratory, Aalto University, P.O. Box
  15100, FI-00076 AALTO, Finland}

\author{M.A. Sillanp\"a\"a}

\affiliation{Low Temperature Laboratory, Aalto University, P.O. Box
  15100, FI-00076 AALTO, Finland}

\affiliation{Department of Applied Physics, Aalto University, P.O. Box 11100, FI-00076 AALTO, Finland }

\date{\today}

\begin{abstract}
Cavity optomechanics is showing promise for studying quantum mechanics
in large systems. However, smallness of the radiation-pressure
coupling is a serious hindrance. Here we show how the charge tuning of the Josephson inductance in a
single-Cooper-pair transistor can be exploited to arrange
a strong radiation pressure -type coupling $g_0$ between mechanical and
microwave resonators. In a certain limit of parameters, such a
coupling can also be seen as a qubit-mediated coupling of two
resonators. We show that this scheme allows reaching extremely high
$g_0$. Contrary to the recent proposals for
exploiting the non-linearity of a large radiation pressure coupling, the
main non-linearity in this setup originates from a cross-Kerr type of
coupling between the resonators, where the cavity refractive index
depends on the phonon number. The presence of this coupling will allow accessing the individual phonon numbers via the measurement of the cavity.
\end{abstract}


\maketitle

Recent experiments on cavity optomechanical systems have shown how the
parametric coupling between an electromagnetic (either optical or
microwave) cavity and a mechanically vibrating resonator can be
exploited to take the latter to its quantum mechanical ground state \cite{teufel11,painter11}. Such schemes rely on amplifying the
intrinsically weak radiation pressure coupling $g_0$ between the two systems
via a strong pumping of the cavity, making the effective coupling
between the systems linear. However, linearly coupled oscillators
constitute a linear system lying in the
correspondence limit, where quantum effects can be seen only in
signal fluctuations \cite{safavinaeni12,massel11}. Therefore, the emphasis of
this research has shifted to the regime of strong radiation pressure
coupling. There are many recent theoretical proposals of the ensuing dynamics
of the system in the strong coupling regime
\cite{mancini97,bose97,ludwig08,nunnenkamp11,rabl11,qian12,xu12,nunnenkamp12,stannigel12,xu13,kronwald13,akram13,liao13},
but reaching this regime in any other system than cold atom gases
\cite{murch08,brennecke08} in practice is challenging \cite{chan12,xuereb12}. The ultimate
aim would be to make the bare radiation pressure coupling of the order
of either the frequency of the mechanical resonator $\omega_m$,  or at
least of the linewidth $\kappa$ of the cavity. 

In this Letter we propose to use the non-linearity of the Josephson
effect to enhance the coupling between the vibrations and the
electromagnetic field. The scheme involves a tripartite system
  consisting of a Josephson junction qubit, a microwave cavity and a
  micromechanical resonator. Although previous work exist on coupling
  a qubit to both a cavity and a mechanical system \cite{Nori06,buks2007,nori2009,pirkkalainen12} our work is to our knowledge the first where the system is considered as an optomechanical platform.

For representative superconducting circuit parameters
\cite{sillanpaa04}, we find that the radiation
pressure coupling can be amplified by a large factor. We first show
this by a simple Josephson inductance picture and then detail a
Schrieffer-Wolff -type approach where the effect is obtained as a
systematic perturbation theory on the tripartite quantum system. Using this
approach we also discuss the possible added mechanical and cavity damping due to
the hybridization of the different parts of the system. Finally, we
continue the perturbation theory to show that the non-linear frequency
shifts in this system are not primarily caused by the radiation pressure
coupling, but rather a cross-Kerr type coupling.

{\em Radiation pressure from Josephson inductance.} Here we take
  advantage of a charge qubit \cite{nakamura,quantronium}, that is, a
  system of two small-capacitance Josephson junctions. This system is
  also known as a single-Cooper-pair transistor (SCPT), which is the
  picture which we first adopt. It behaves as a tunable inductance dependent on the mechanical displacement.

\begin{figure}[h]
\centering
\includegraphics[width=\columnwidth]{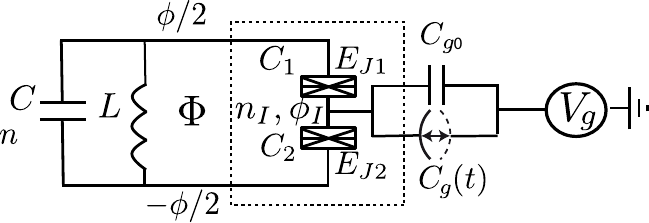}
\caption{Microwave optomechanical circuit considered here. The SCPT part is marked with a dashed box. The mechanical resonator couples via a time-dependent capacitance $C_g[x(t)]$.}
\label{fig:setup}
\end{figure}

As marked by the dashed box in Fig.~1, the SCPT has the junction
capacitances $C_1, C_2$, and the gate capacitance $C_{g0}$ which give
the charging energy of a single electron $E_C = e^2/[2
(C_{g0}+C_1+C_2)]$. The junctions have the Josephson energies $E_{Ji}
\lesssim E_C$. Due to Coulomb blockade, the energy difference of
having zero or one Cooper pairs on the island can be tuned by the gate
charge $n_{g0}=V_g C_{g0}/2e$. Moreover, Josephson tunneling mixes
charge states into coherent superpositions of Cooper pair states \cite{makhlin01}. 

In the most relevant limit $E_C \gg E_{Ji}$, so that we concentrate on the two charge states closest to $n_{g0}$, defining $\delta n_{g0}=n_{g0}-{\rm  int}(n_{g0}) \in [0,1]$ as the deviation of $n_{g0}$ from the lower integer value. The Hamiltonian is \cite{dimensionnote}
\begin{equation}
H_{\rm SCPT}=\sum_{j=1}^3 B_j \sigma_j/2,
\label{eq:chargestateHamiltonian}
\end{equation}
where $B_1=-(E_{J1}+E_{J2}) \cos(\phi/2)$, $B_2=(E_{J1}-E_{J2})\sin(\phi/2)$ and
$B_3=4 E_C(1-2 \delta n_{g0})$ are the effective magnetic fields,
and $\sigma_j$ are Pauli matrices acting on the space spanned by the Cooper-pair charge states
$|{\rm int}(n_g)\rangle$ and $|{\rm int} (n_g) + 1\rangle$, and $\phi$ is the phase difference of the superconducting order parameters across the
junction. The ground state energy is $E_{\rm SCPT}=-\sqrt{\sum_j  B_j^2}/2 \equiv -B/2$. 

Placing a Josephson
junction inside an electromagnetic resonator (cavity) affects its total
inductance via the Josephson inductance $L_J=\hbar^2/(2e)^2[\partial_\phi^2
E(\phi)]^{-1}$, where $E(\phi)$ is the energy of the junction. Using an SCPT instead of a single junction allows for controlling Josephson inductance via the modulation of the gate charge \cite{sillanpaa04}. Here
we consider what happens when the gate capacitor can vibrate,
modulating the movable part of the gate capacitance $C_g(x,t)$ (see Fig.~\ref{fig:setup}). The total gate charge is $n_g=  n_{g0}+x V_g \partial_x C_g$, where $x$ is the amplitude of
mechanical vibrations.  Along the dependence of the energy of the SCPT
on both control parameters $n_g$ and $\phi$, the mechanical vibrations modulate the cavity eigenfrequency, and the resulting coupling is of the radiation pressure type.

The above picture allows us to estimate the size of the radiation
pressure coupling. The cavity eigenfrequency $\omega_c=[(L || L_J)
C]^{-1/2}$ consists of the geometric and the Josephson inductances $L$
and $L_J$, respectively. The radiation pressure coupling is thus
\begin{equation}
\label{eq:generalradpressure}
g_0\equiv x_{\rm {ZP}}\frac{\partial \omega_c}{\partial
  x}=\frac{\omega_c x_{\rm{ZP}} \partial_x C_g V_g}{4e} \frac{LL_J}{L+L_J} \partial_{n_g} L_J^{-1}.
\end{equation}
Here $x_{\rm{ZP}}=\sqrt{\hbar/(2m \omega_m)}$ is the 
zero-point motion amplitude for a mechanical resonator with effective mass $m$ and angular frequency
$\omega_m$. Let us compare this with the coupling in the setup where the capacitance
of the cavity is directly modulated \cite{regal08,massel11,teufel11}. In that case
$g_0^d=x_{\rm{ZP}} \partial_x C_g/(2 C) \omega_c$. The ratio between
these two couplings is
\begin{equation}
\label{eq:ratioofcouplings}
\frac{g_0}{g_0^d} = \frac{C V_g}{2e} \frac{L
  L_J}{L+L_J} \partial_{n_g} L_J^{-1}.
\end{equation}
Choosing $L \approx L_J$ and noting that the factor
$L_J \partial_{n_g} L_J^{-1}$ can be of the order of unity (see below), the
optomechanical coupling can be amplified in this setting by the factor
$C V_g/(2e)$, about 4 to 6 orders of magnitude for typical experimental parameters  \cite{sillanpaa04,lahaye09}.  

\begin{figure}[h]
\centering
\includegraphics[width=\columnwidth]{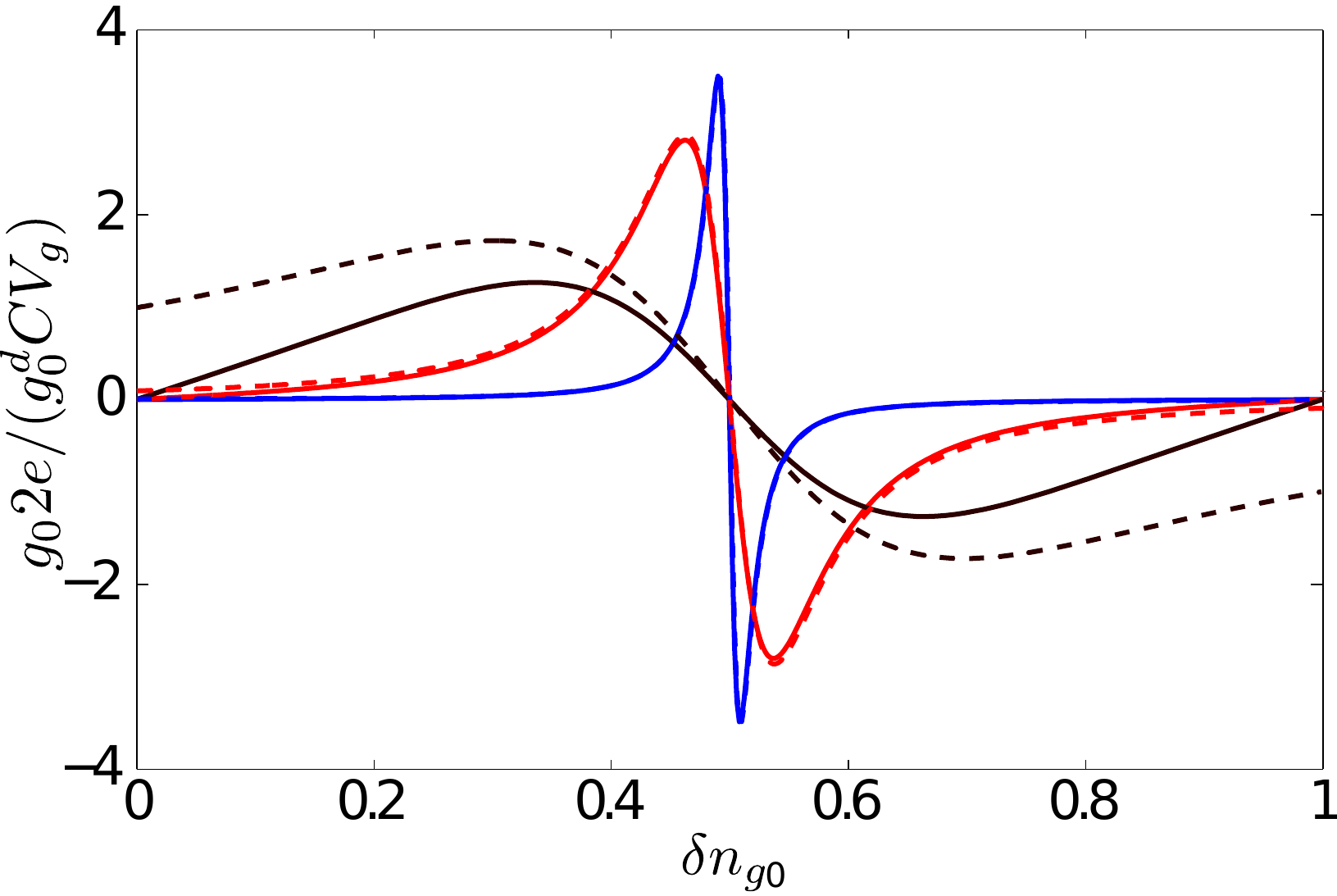}
\caption{(Color online): Radiation pressure coupling $g_0$ vs. gate charge $\delta
  n_{g0}$ for the case of equal Josephson couplings ($E_{J1}=E_{J2}$),
  at flux $\Phi=0$ and for three different ratios $E_J/E_C=0.05$
  (blue), 0.2 (red) and 1 (black). The inductance $L=20
  \hbar^2/(4e^2 E_C)$ in each curve. Solid lines show the results from
  the numerically obtained SCPT spectrum beyond the two-charge state restriction in \eref{eq:chargestateHamiltonian}, whereas the dashed lines are plotted from
  Eq.~\eqref{eq:radpressure}.}
\label{fig:rpcoupling}
\end{figure}

The radiation pressure coupling is now straigthforward to obtain from
Eq.~\eqref{eq:ratioofcouplings}. For symmetric junctions,
$E_J=E_{J1}=E_{J2}$, we get
\begin{equation}
\label{eq:radpressure}
\frac{g_{0}}{g_{0}^d}=\frac{C V_g}{2e} \frac{8 E_C^2 E_J^2 l \left(1-2 \delta
      n_{g0}\right)}{\left(4 \tilde E_C^2
   +E_J^2\right) \left(4
   \sqrt{E_J^2+4 \tilde E_C^2 }+E_J^2
   l\right)},
\end{equation}
where $\tilde E_C=E_C(1-2 \delta n_{g0})$ and $l=L (2e)^2/\hbar^2$. We
plot $g_0$ vs.~$\delta n_{g0}$ in Fig.~\ref{fig:rpcoupling}. The
two-state approximation is generally valid for $E_C \gtrsim E_J$ as
long as $n_{g0}$ is not too close to an integer. For low $E_J/E_C$, $g_0$ contains a peak of width $\sim
E_J/E_C$ with a maximum somewhat below the charge
degeneracy point $\delta n_{g0}=1/2$. For $l = 1/E_J$, the maximum
resides at $\delta n_{g0} \approx 1/2-0.18 E_J/E_C$ and is $\max_{\delta
  n_{g0}} g_0 \approx 0.32 \, g_{0}^d C V_g/(2e) E_C/E_J$.  The largest
$g_0$ is thus obtained in the extreme charge qubit
limit $E_C \gg E_J$, but because the range of gate charge values where
this maximum is obtained is proportional to $E_J/E_C$, in practice it
is preferable to choose $E_J$ not too far from $E_C$ to prevent gate
charge fluctuations from masking the effect.

To make a numerical estimate of the resulting radiation pressure
coupling, let us choose $(2e)^2 L/\hbar^2 = E_J=E_C/2$. For some representative values $C=50$ fF, $V_g=10$ V, we would get $g_0/g_0^d \approx
10^6$. With the typical direct radiation pressure coupling $g_0^d/(2 \pi) \sim 10$
Hz \cite{sulkko10}, we would hence get $g_0/(2 \pi) \sim 10$ MHz, which
is already of the order of typical $\omega_m$ and two orders of
magnitude larger than a typical Al cavity linewidth $\kappa$
\cite{teufel11} below 100 mK. However, in practice the linewidths are affected by charge noise of the qubit (see below).

\begin{figure}[t]
\centering
\includegraphics[width=\columnwidth]{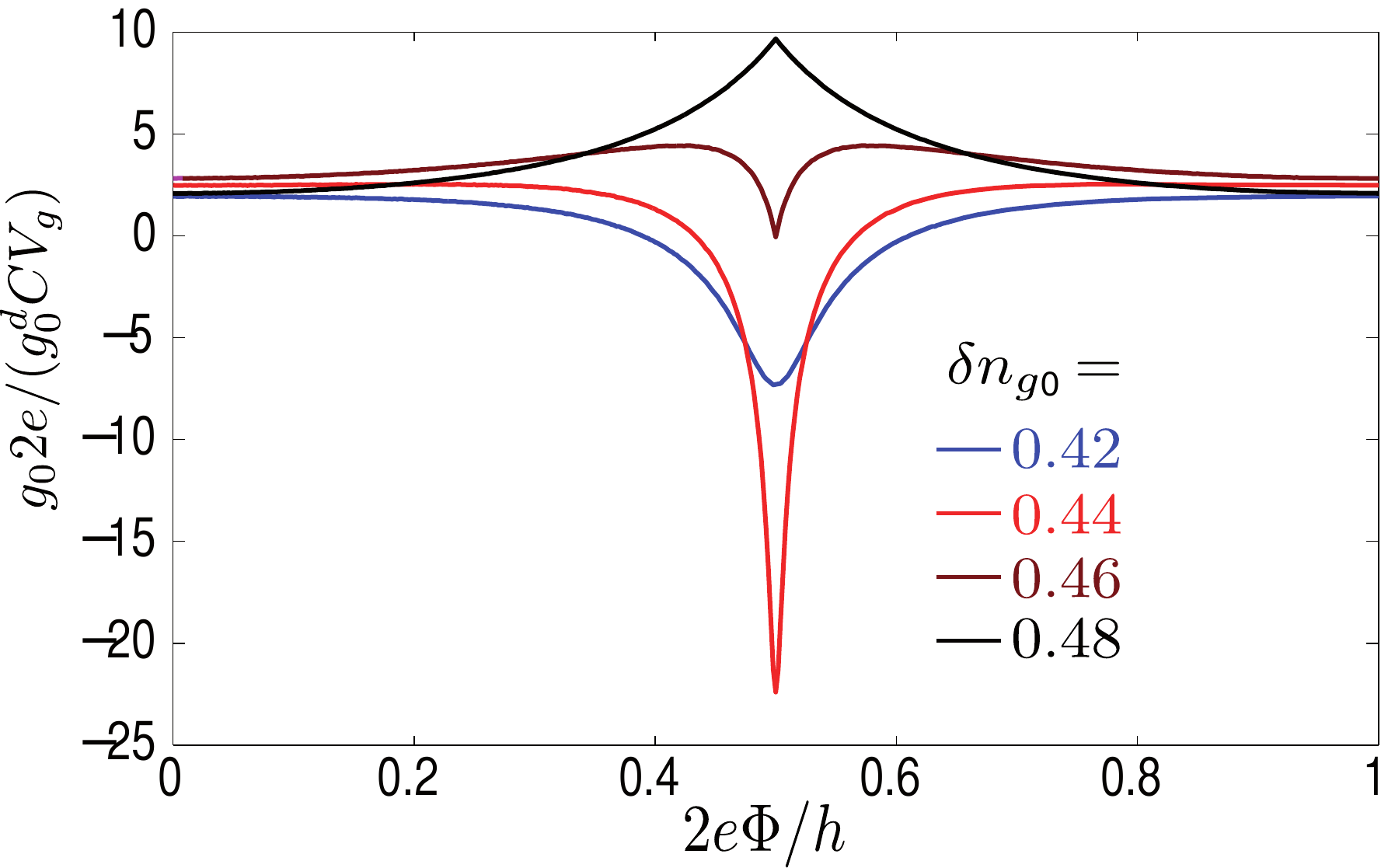}
\caption{(Color online): Radiation pressure coupling $g_0$ vs.~flux bias
  for a few values of the gate charge indicated in the plot for
  $E_{J1}=E_{J2}=0.2E_C$ and $L=20 \hbar^2/(4e^2
  E_C)$.}
\label{fig:fluxdependence}
\end{figure}

In the presence of a flux $\Phi$ through the cavity loop, the average phase
$\phi_a=\langle \phi \rangle$ is the phase that minimizes the total energy \cite{minimumnote} 
\begin{equation}
\label{eq:phaseeq}
\frac{\hbar^2}{8e^2 L} \left(\frac{2e\Phi}{\hbar}-\phi\right)^2 + E_{\rm{SCPT}}(\phi, n_{g0}).
\end{equation}
For a vanishing flux, $\phi_a=0$. It is then possible to tune the radiation pressure coupling with the flux, see Fig.~\ref{fig:fluxdependence}. Interestingly, such a flux tuning is stronger for smaller $E_J$, but the region of gate charges where $g_0$
is appreciable is again limited to a range proportional to
$E_J/E_C$. \

Another way to see why the coupling is boosted is because of the
  qubit nonlinearity. A mechanical resonator can be coupled to a
  linear cavity (not to a qubit as in the present work) by means of a
  voltage bias \cite{tian}. This coupling has a magnitude comparable
  to $g_0$. This coupling, however, is linear, and has little
  consequences between two linear resonators. Replacing the cavity
  by a qubit, however, turns the linear coupling into a longitudinal
  coupling, which has a strong influence on the energies.

{\em Schrieffer-Wolff approach.} The Josephson inductance approach provides an intuitive picture of the physics. However, for a more rigorous treatment, we start from the general tripartite
Hamiltonian and derive the optomechanical coupling by using the
Schrieffer-Wolff transformation \cite{bravyi11}. It consists in a unitary transformation which uncouples the high- and the
low-energy states, leading to the definition of an effective
low-energy Hamiltonian. Here, the high-energy states are
represented by the qubit states while the low-energy ones are
represented by cavity and mechanical oscillator modes. The ensuing
effective low-energy Hamiltonian is thus described in terms  of effective
cavity and mechanical oscillator modes. Note that both this and the
above Josephson inductance approach are valid only in the dispersive
limit, where $\hbar \omega_{c/m} \ll |B|$.

We express the electromagnetic energies of the circuit in Fig.~\ref{fig:setup} in terms of the phases $\phi$
and $\phi_I$ with conjugate charges $2en$ and $2en_I$ ($n$ and $n_I$
denote the number of Cooper pairs, and $I$ points to the SCPT
  island). Then we use the fact that $e^{i\phi_I}$ is a ladder
operator for charge $n_I$ \cite{makhlin01}, assume that changes of
$\phi$ with respect to $\phi_a$ are small compared to $2\pi$ (in the opposite limit the
dynamics of the system is quite complicated, see for example
\cite{gramich13}) and define $\phi-\phi_a=\phi_0 (c^\dagger+c)$ with
the conjugate variable $n=i \sqrt{\hbar/(8e^2
  Z_0)}(c^\dagger-c)$. Here, $Z_0=\sqrt{L/C}$ and the phase
  zero-point fluctuation is $\phi_0=\sqrt{2e^2 Z_0/\hbar}$. We then
write the resulting Hamiltonian for the system in
the charge basis as above. 
With a
similar quantization of the mechanical part of the Hamiltonian, we get
(see detailed derivation in the Appendix)
\begin{equation}
\begin{split}
&H=H_{\rm SCPT} + \hbar \omega_c^0 c^\dagger c + \hbar
\omega_m^0 a^\dagger a  + g_m \sigma_3
(a^\dagger+a) + \\&
(g_{q1} \sigma_1 +g_{q2} \sigma_2) (c^\dagger+c)^2+(g_{c1}\sigma_1+g_{c2}\sigma_2)(c^\dagger+c).
\end{split}
\end{equation}
Here $\omega_c^0=(LC)^{-1/2}$ and $\omega_m^0$ are the eigenfrequency of the bare $LC$
oscillator and of the bare mechanics, respectively, $B_j$ are as above and
\begin{subequations}
\begin{align}
g_m&=-\frac{4 E_C x_{\rm{ZP}} \partial_x C_g
V_g}{2 e}\\ 
g_{q1}&=\frac{e^2 Z_0}{8\hbar} (E_{J1}+E_{J2}) \cos(\phi_a/2)\\
g_{q2}&=\frac{e^2 Z_0}{8\hbar} (E_{J2}-E_{J1}) \sin(\phi_a/2)\\ 
g_{c1}&=\sqrt{\frac{e^2 Z_0}{8\hbar}} (E_{J1}+E_{J2}) \sin(\phi_a/2)\\ 
g_{c2}&=\sqrt{\frac{e^2 Z_0}{8\hbar}} (E_{J1}-E_{J2}) \cos(\phi_a/2).
\end{align}
\end{subequations}
The system is thus composed of two resonators with frequencies
$\omega_c^0$ and $\omega_m^0$ coupled to a common qubit. We note that the
order of magnitude of the cavity couplings satisfy $g_{qj} \sim
g_{cj}^2/E_J \gg g_m$. Below we limit
ourselves to the case of a symmetric system, $E_{J1}=E_{J2}$ in which
case $B_2=g_{q2}=g_{c2}=0$. The full results are given in
the Appendix. 

The full Schrieffer-Wolff transformation diagonalizing the qubit part
of the system is quite complicated. However, in the dispersive limit
$\hbar\omega_{c,m} \ll |B|$ and where all couplings are smaller than
the difference $|B|-\hbar \omega_{c,m}$, it is enough to diagonalize
the qubit treating the oscillator coordinates as scalars. 
Assuming that this effective qubit stays in its ground state,
we may replace
$\sigma_3 \rightarrow -1$. Expanding in the coupling constants,
we get a preliminary form of the Hamiltonian:
\begin{equation}
\label{eq:effham}
\begin{split}
H_{\rm eff}=&\hbar \omega_c^0 c^\dagger c + \hbar \omega_m^0 a^\dagger a
+\alpha_c \hat x_c+\alpha_m \hat x_m+\\&\hbar g_{cm}
\hat x_c\hat x_m+\hbar g_{Sc} \hat x_c^2 + \hbar g_{Sm}
\hat x_m^2 + \hbar g_{rp} \hat x_c^2\hat x_m.
\end{split}
\end{equation}
Here $\hat x_c \equiv (c^\dagger+c), \hat x_m \equiv (a^\dagger+a)$ and the
coefficients are $\alpha_c=-B_1 g_{c1}/B$, $\alpha_m=-B_3 g_m/B$,
$\hbar g_{cm}=2B_3 B_1 g_{c1} g_m/B^3$ and
\begin{equation*}
\begin{split}
\hbar g_{Sm}&=-\frac{B_1^2 g_m^2}{B^3}, \quad
\hbar g_{Sc}= -\frac{B_3^2 g_{c1}^2-B^2 B_1 g_{q1}}{B^3}\\
\hbar g_{rp} &= \frac{2B_3 g_m \left(B^2 B_1
    g_{q1}+(B_3^2-2B_1^2) g_{c1}^2\right)}{B^5}.
\end{split}
\end{equation*}
The first two terms in Eq.~\eqref{eq:effham} are the Hamiltonians for
the bare oscillators, the next two are qubit-induced static forces on
them (neglected below), the term proportional to $g_{cm}$ is a linear coupling between
the oscillators, and the terms with coefficients $g_{Sc}$ and $g_{Sm}$
are the cavity and mechanical Stark shifts \cite{schuster05,lahaye09,pirkkalainen12}. Finally, $g_{rp}$ denotes an intermediate expression of a radiation-pressure type coupling. 
We diagonalize the individual oscillator Hamiltonians by the Bogoliubov
transformation, introducing $c=\cosh(\theta_{c}) d + \sinh(\theta_{c})
d^\dagger$ and $a=\cosh(\theta_{m}) b + \sinh(\theta_{m})
b^\dagger$ with $\theta_{c/m}=-{\rm
  atanh}[(2g_{Sc/m})/(\hbar\omega_{c/m}^0+2g_{Sc/m})]/2$. This yields
the effective frequencies $\omega_c=\sqrt{\omega_c^0(\omega_c^0+4
  g_{Sc})}$ and $\omega_{m}=\sqrt{\omega_{m}^0(\omega_{m}^0+4 g_{Sm})}$ for the
cavity and the vibrations, respectively. The linear coupling $g_{cm}$ has little
effect, and is neglected. Including the Bogoliubov
transformations into the radiation pressure coupling and assuming
$g_{rp}x_m  \ll \omega_c$ yields a coupling between the effective mechanics and cavity of the form $\hbar g_{0} d^\dagger d (b^\dagger+b)$.
Here $g_0=2 g_{rp}(\omega_c^0/\omega_c)(\omega_m^0/\omega_m)^{1/2}$
\cite{bogoliubovnote}  is the radiation pressure
coupling. These results then coincide with those of the above Josephson
inductance approach, except for the renormalization due to
mechanical Stark shift $(\omega_m^0/\omega_m)^{1/2}$ that is not
captured by the latter. 

{\em Quantum non-linearities.} The possibility of obtaining a large Josephson-enhanced radiation
pressure coupling $g_{0}$ implies a good prospect of reaching the ``quantum
regime'' of optomechanics, where $g_{0}$ becomes at least of the order of the
cavity linewidth $\kappa$. In this regime it should be possible to observe non-linearities directly in the spectrum.
The frequency shift is proportional to $g_{0}^2/\omega_m$
\cite{rabl11}, and is of the
order $\sim g_{cj}^4 g_m^2 \sim g_{qj}^2 g_m^2$. However, the qubit-mediated coupling has
another non-linearity that gives rise to a frequency shift in the
mechanics and shows
up at a lower order. It can be understood as the change in the photon Stark
shift (which depends on the qubit level splitting) due to the
phonon-driven qubit Stark shift, and it implies a term of the form
$c^\dagger c a^\dagger a$. It is thus of the form of the cross-Kerr
effect between the two resonators. In the perturbation series with respect to
the couplings $g_{cj}, g_{qj}$ and $g_m$, such a term would be of the order
$g_{cj}^2 g_m^2 \sim g_{qj} g_m^2$.
In the rotating wave approximation we get the term $H_{\m{cK}}=\hbar g_{\m{cK}} d^\dagger d b^\dagger b$ with
\begin{equation}
\label{eq:gck}
\frac{g_{\m{cK}}}{g_0}=\frac{R_Q}{Z_0}\frac{\hbar g_0}{E_J}\frac{ \left(4
    \tilde E_c^2 \left(8 E_s+E_J^2 l\right)-E_J^2 \left(4 E_s+E_J^2
      l\right)\right)}{2 \pi  \tilde E_c^2 E_J\sqrt{4+\frac{E_J^2 l}{E_s}}},
\end{equation}
where $R_Q=h/(2e)^2$ and $E_s=\sqrt{4 \tilde E_C^2+E_J^2}$. The total effective optomechanical Hamiltonian is thus 
\begin{equation}
\label{eq:totham}
H_{\rm eff}=\hbar \omega_c d^\dagger d + \hbar \omega_m b^\dagger b + \hbar g_{0} d^\dagger d (b^\dagger+b) + H_{\m{cK}}.
\end{equation}
Whereas the response with side-band
driving is dominated by the large $g_{0}$, the nonlinear frequency
shifts are mainly due to the term $g_{\m{cK}}$ (see the spectrum of
Eq.~\eqref{eq:totham} in the Appendix). For example, for $1/l \approx
E_J \approx \tilde E_c$, where the radiation pressure coupling $g_0$
is appreciable, we get $g_{\m{cK}}/g_0 \approx 5 \, \hbar g_0 R_Q/(E_J
Z_0)$. With $g_0 \approx 5$ MHz, $E_J/\hbar\approx 10$ GHz,
and $Z_0 \approx R_Q/100$, we would hence get an appreciable
nonlinearity, $g_{\m{cK}} \approx 0.25 \, g_0$. Moreover, close to $n_{g0}=1/2$, the radiation pressure term vanishes whereas the cross-Kerr term is finite. The Hamiltonian becomes particularly simple, as the coupling commutes with the rest of the Hamiltonian. As a result, the cavity frequency is shifted by the number of quanta in the mechanical resonator. Such a shift could be used for a direct detection or creation of the Fock states in the mechanical resonator.

{\em Effect of qubit-mediated dissipation.} Since the qubit and the
oscillators are generally hybridized up to a significant amount, it is
important to consider the effect of qubit energy relaxation on
that of the oscillators. 
This can be analyzed with the Schrieffer-Wolff
approach, but now applying the transformation only to the qubit-oscillator
part of the setup (see also the Appendix). 
We find that the rates for relaxation or excitation of the mechanical resonator due to the qubit
dissipation satisfy 
\begin{equation}
\frac{\gamma_{\rm rel/exc}}{\gamma_{\rm rel/exc}^q} = \frac{2 g_{m}^2 B_{1}^2}{B^4} \frac{\sum_j
  \lambda_j^2 S_j(\pm
  \omega_{m})}{\sum_j \lambda_j^2 S_j(\pm B)}.
\label{eq:dampingbyhybridization}
\end{equation}
Here $\gamma_{\rm rel,exc}^q$ are the bare qubit relaxation/excitation
rates, $\lambda_j$ coupling between the qubit and the bath oscillator $j$, and $S_{j}(\omega) \equiv \int
dt e^{i\omega t} \langle [b_j^\dagger(t)+b_j(t)]
[b_j^\dagger(0)+b_j(0)]\rangle$ is the correlator of the qubit bath,
chosen diagonal for convenience. For an equilibrium bath at
temperature $T_b$, these satisfy a detailed balance relation
$\gamma_{\rm rel}/\gamma_{\rm
  exc}=S_{j}(\omega_{m})/S_{j}(-\omega_{m})=\exp[\hbar \omega_{m}/(k_B
T_b)]$. Using the fluctuation-dissipation relation with a frequency
independent susceptibility for the bath correlator (i.e., quantum
noise increasing linearly with an increasing frequency), we would then get
at $k_B T_b \lesssim \hbar \omega_{m}$ the induced mechanical dissipation rate
$\gamma_{\rm rel} \approx g_{m}^2 \omega_{m} B_{1}^2 \gamma_{{\rm
    rel}}^q/B^5 $, which is likely quite small in practical
systems. However, in the case of charge qubits, one should consider 
$1/f$ (flicker) noise, i.e., noise increasing linearly with a
decreasing frequency. In this case this relation changes
roughly to $\gamma_{\rm
  rel} \approx g_{m}^2B_{1}^2/(B^3 \omega_{m}) \gamma_{q, {\rm rel}}$, which may
already become relevant compared to the intrinsic oscillator
dissipation at large values of $g_{m}$. 

In the case of the cavity, we get similar effects on relaxation and
excitation rates by above replacing $g_m$ with $g_{c1}$ or $g_{c2}$,
$B_1$ by $B_3$, and $\omega_m$ by $\omega_c$. However, a more relevant
effect is likely due to pure cavity dephasing seen by a flickering of
the cavity frequency due to low-frequency background charge
fluctuations in the qubit. As analyzed in the Appendix, the rate for
this process in the case $E_{J1}=E_{J2}$ and $\phi_a=0$ is $\gamma_\phi \sim
g_{q1}^2 \gamma_\phi^q/B^2$, where $\gamma_\phi^q$ is the pure
dephasing rate of the qubit. At the optimal operation point, this is
of the order of $e^4 Z_0^2 \gamma_\phi^q/(16\hbar^2)$. For example,
using $\gamma_\phi^q=500$ MHz
\cite{gunnarsson08} and $Z_0 \approx 500$ $\Omega$, the corresponding added cavity dephasing rate
would be of the order of 0.5 MHz, which is already larger than the
intrinsic $\kappa$. Nevertheless, as $g_0/\gamma_{\phi} \propto V_g$, this does
not hinder reaching the limit of strong optomechanical coupling.

Note that the above approach is valid as long as the cavity photon
number $n_c$ is not too large, such that phase fluctuations $\phi_0 \sqrt{n_c}$
  are small compared to $2 \pi$. Typical values are $\phi_0
  \sim 0.1...0.5$, depending on the cavity impedance. At the upper end
  of the scale there are nonlinear corrections to energy of 10 \%
  already at $n_c \sim 1$. In spite of the low  linear regime, because
  of the large $g_0$, optomechanical phenomena are overwhelming
  already at photon numbers $n_c \approx 1$.

Summarizing, we have presented a realizable scheme for boosting the optomechanical radiation pressure coupling by several orders of magnitude. This gives the possibility to approach the previously elusive single-photon strong coupling limit of optomechanics. Our predictions can be readily tested in the state of the art circuit optomechanical devices.

We acknowledge fruitful discussions with Pertti Hakonen, Florian
Marquardt and Sorin Paraoanu. This work
was supported by the European Research Council (Grants
No.~240362-Heattronics, 240387-NEMSQED and MPOES) and the EU-FP 7 INFERNOS
(Grant No. 308850) and MICROKELVIN (Grant No. 228464) programs.

\appendix
\begin{widetext}

{\bf APPENDIX}

\section{Derivation of Eq. (6)}

We start by writing the electromagnetic energy of the tripartite quantum system:
\begin{equation}
\begin{split}
H &= 4 E_{Cc}
n^2+4 E_C(n_I-n_{g}(x))^2 + \frac{p^2}{2m}+\frac{\hbar^2}{4e^2}\frac{(\phi-2e\Phi/\hbar)^2}{2L}\\&-E_{J1}
\cos(\phi/2-\phi_I)-E_{J2} \cos(\phi/2+\phi_I) + \frac12 m\omega_m^0 x^2,
\end{split}
\end{equation}
where we have allowed also for the mechanical modulations of the gate charge $n_g(x)=n_{g0}+\frac{1}{2e}x\partial_x C_gV_g$, and introduced the flux $\Phi$ through the cavity loop. We denote with $\phi_a=\langle \phi\rangle$ the average phase, i.e., the phase that minimizes the total energy. For a vanishing flux, $\phi_a=0$. We quantize the cavity and the mechanical oscillator by defining the annihilation operators
\begin{equation}
\begin{split}
c&=\frac{1}{\sqrt{2\hbar C\omega_c^0}}\bigg[C\omega_c^0\bigg(\frac{\hbar}{2e}(\phi-\phi_a)\bigg) +i 2en\bigg]\\
a&=\frac{1}{\sqrt{2\hbar m\omega_m^0}}(m\omega_m^0x +i p),
\end{split}
\end{equation}
of single quanta in the cavity and the oscillator, respectively. The conjugate variables can now be written as
\begin{equation}
\begin{split}
\phi-\phi_a &= \sqrt{\frac{2e^2Z_0}{\hbar}}(c^{\dag}+c), \ \ n=i\sqrt{\frac{\hbar}{8e^2Z_0}}(c^{\dag}-c),\\
x &= x_{ZP}(a^{\dag}+a), \ \ p=i\sqrt{\frac{\hbar m\omega_m^0}{2}}(a^{\dag}-a),
\end{split}
\end{equation}
where $x_{ZP}=\sqrt{\hbar/(2m\omega_m^0)}$. We end up with
\begin{equation}
\begin{split}
H &= \hbar\omega_c^0 c^{\dag} c + \hbar\omega_m^0 a^{\dag} a+4 E_C(n_I-n_{g}(x))^2 \\&-(E_{J1}+E_{J2})\cos\big[\eta(c^{\dag}+c)+\phi_a/2\big]\cos \phi_I - (E_{J1}-E_{J2})\sin\big[\eta(c^{\dag}+c)+\phi_a/2\big]\sin \phi_I,
\end{split}
\end{equation}
where $\eta=\sqrt{e^2Z_0/(2\hbar)}$ and we have expanded the Josephson potentials using trigonometric identities.

In the most relevant limit $E_C \gg E_{Ji}$, and it is enough to concentrate on the two charge states closest to $n_{g0}$, defining $\delta n_{g0}=n_{g0}-{\rm  int}(n_{g0}) \in [0,1]$ as the deviation of $n_{g0}$ from the lower integer value. This results in the replacements
\begin{equation}
\begin{split}
\cos \phi_I &\rightarrow \frac12 \sigma_1, \ \ \sin \phi_I \rightarrow -\frac12 \sigma_2, \ \ (n_I-n_g(x))^2\rightarrow \frac12(1-2\delta n_g(x))\sigma_3.
\end{split}
\end{equation}
In this approximation, the Hamiltonian can be written in the form
\begin{equation}
\begin{split}
H &= \hbar\omega_c^0 c^{\dag} c + \hbar\omega_m^0 a^{\dag} a  + 2 E_C(1-2 \delta n_{g0}) \sigma_3 - g_m\sigma_3(a^{\dag}+a)\\&-\frac12\cos[\eta(c^{\dag}+c)]\bigg[(E_{J1}+E_{J2})\cos\frac{\phi_a}{2}\sigma_1 - (E_{J1}-E_{J2})\sin\frac{\phi_a}{2}\sigma_2\bigg]\\
&+\frac12\sin[\eta(c^{\dag}+c)]\bigg[(E_{J1}+E_{J2})\sin\frac{\phi_a}{2}\sigma_1 + (E_{J1}-E_{J2})\cos\frac{\phi_a}{2}\sigma_2\bigg]
\end{split}
\end{equation}
where $g_m=4E_C x_{ZP}\partial_x C_gV_g/(2e)$ and we again applied trigonometrics. By writing this in second order in $\eta(c^{\dag}+c)$, we obtain Eq.(7) of the main text.

\section{General coefficients of Hamiltonian (8)}
In the main text we assume the case of symmetric junctions, $E_{J1}=E_{J2}$ in order to keep the formulas short. In the general case the coefficients of Hamiltonian (8) are
\begin{equation*}
\begin{split}
\alpha_c &= -\frac{B_1 g_{c1}+B_2 g_{c2}}{B}, \quad
\alpha_m=-\frac{B_3 g_m}{B}, \quad 
\hbar g_{cm}=\frac{2B_3(B_1 g_{c1}+B_2 g_{c2}) g_m}{B^3}, \quad
\hbar g_{Sm}=-\frac{(B_1^2+B_2^2) g_m^2}{B^3}\\
\hbar g_{Sc}&= \frac{(B_1 g_{c1}+B_2
  g_{c2})^2-B^2(g_{c1}^2+g_{c2}^2+B_1 g_{q1}+B_2 g_{q2})}{B^3}\\
\hbar g_{rp} &= \frac{2B_3 g_m \left(B^2 (g_{c1}^2+g_{c2}^2+B_1
    g_{q1}+B_2 g_{q2})-3(B_1 g_{c1}+B_2 g_{c2})^2\right)}{B^5}.
\end{split}
\end{equation*}
The fields $B_j$ are given below Eq.~(4) and the resonator-qubit coefficients in Eq.~(7) of the main text.

\section{Damping by hybridization}

We study the damping induced by the qubit on an effective oscillator coordinate by exploiting the Schrieffer-Wolff
approach, but now applying the transformation to the qubit-mechanics
or the qubit-cavity
part of the setup. To investigate the influence of the qubit energy
relaxation on the linewidth of the effective resonator coordinate,
taking into account the hybridization, we start from the description
of this part of the system, written in the diagonal basis of the
qubit ($\hbar=k_B=1$ in this discussion),
\begin{equation}
\label{eq:ham}
\begin{split}
H=\underbrace{\frac{B}{2} \sigma_3+\sum_j \bigg[\omega_j b_j^\dagger
  b_j}_{H_0} + \underbrace{\lambda_j (b_j^\dagger+b_j) \sigma_1}_{V}\bigg]
+\omega_m
a^\dagger a + g_m (a^\dagger + a)\sigma_1.
\end{split}
\end{equation}
For simplicity, we only consider the mechanical resonator and we drop
out the terms parallel with the qubit Hamiltonian in the
qubit-resonator coupling, as they do not directly contribute to the
hybridization. We also ignore the bath of the oscillator, as we are interested in the added dissipation due to the qubit-oscillator hybridization. To connect with the results of the main paper, we hence
need to replace $g_m \mapsto g_m B_1/B$. This calculation also
describes the damping effect in the cavity due to the cavity couplings
$g_{c1}$ and $g_{c2}$ --- the other coupling terms are considered below. In Eq.~\eqref{eq:ham} the second term is the Hamiltonian of the qubit bath, and the
third term the coupling between the qubit and the bath.  

Consider first the derivation of the standard Lindblad equation for a qubit. We hence first take $g_m=0$.
Now let us consider the coupling $V$ between the systems as a perturbation. Ignoring the frequency shifts due to the bath coupling, we get in the second order in perturbation theory the usual equation for the qubit density matrix 
\begin{equation}
\dot \rho_q = i[\rho_q,H_0]-\int_{t_0}^t dt_1 {\rm Tr_R} [V(t),[V(t_1) \rho_R \rho_q(t)]],
\end{equation}
where $\rho_R$ is the reservoir density matrix and $t_0$ some initial time of evolution. In the following, we make the (Markovian) approximation that the reservoir stays in the thermal state regardless of the interaction. Above, 
\begin{equation}
V(t)= \exp(iH_0 t) V \exp(-iH_0 t)=(\sigma_\uparrow e^{iB t}+\sigma_\downarrow e^{-iB t})\sum_j \lambda_j x_j(t) \equiv \Gamma(t) \sum_j \lambda_j x_j(t).
\label{eq:timedepV}
\end{equation}
We introduce the correlator 
\begin{equation}
S_{jk}(\omega)=\int dt e^{i\omega t} {\rm Tr}_R \rho_R x_j(t) x_k(0).
\end{equation}
We can assume for simplicity and without loss of generality that $S_{jk}(\omega) = S_j(\omega) \delta_{jk}$ is diagonal. After a few straightforward steps we then obtain
\begin{equation}
\begin{split}
\dot \rho_q = i[\rho_q,H]-\sum_j \lambda_j^2 \int \frac{d\omega}{2\pi} \int_{t_0}^t dt' e^{-i\omega t'} \bigg[&S_j(\omega) [\Gamma(t) \Gamma(t_1) \rho_q(t)-\Gamma(t_1) \rho_q(t) \Gamma(t)]\\&S_j(-\omega)[\rho_q(t) \Gamma(t_1) \Gamma(t)-\Gamma(t) \rho_q(t) \Gamma(t_1)]\bigg].
\end{split}
\end{equation}
Here $t'=t-t_1$. Now, we substitute Eq.~\eqref{eq:timedepV} for $\Gamma(t)$, and disregard terms of the form $\sigma_\uparrow \rho_q \sigma_\uparrow$ and $\sigma_\downarrow \rho_q \sigma_\downarrow$ since such terms contain an explicit oscillatory dependence on time $t$ and therefore their contribution vanishes in the long time limit. We get integrals of the form
\begin{equation}
\int \frac{d\omega}{2\pi} S(\omega) \int_{t_0}^t dt' e^{i(B-\omega) t'}.
\end{equation}
Assuming a long time has passed since the initial time, we can replace the limits by infinities. Therefore, the above integral would yield simply $S_j(B)$. Now it is straightforward to write the resulting qubit dissipator
\begin{equation}
{\cal L}_q=\sum_j \lambda_j^2 \left[S_j(B) \left(\sigma_\downarrow \rho_q(t) \sigma_\uparrow -\{\rho_q(t),\sigma_\uparrow \sigma_\downarrow\}\right) + S_j(-B) \left(\sigma_\uparrow \rho_q(t) \sigma_\downarrow-\{\rho_q(t),\sigma_\downarrow \sigma_\uparrow\}\right)\right].
\end{equation}
We can thus identify the qubit relaxation and excitation rates
\begin{equation}
\gamma^q_{\rm rel}=2\sum_j \lambda_j S_j(B), \quad  \gamma^q_{\rm exc}=2\sum_j \lambda_j S_j(-B).
\end{equation}

Now let us consider a strongly coupled qubit-oscillator system, described by the Hamiltonian in Eq.~\eqref{eq:ham}.
The two systems can be decoupled utilizing the Schrieffer-Wolff transformation 
\begin{equation}
U_{SW}=\exp(S),
\end{equation}
where 
\begin{equation}
S=\frac{g_m}{\omega_m^2-B^2}\left[\omega_m\sigma_1 -iB \sigma_2(a^\dagger+a)\right]  + o(g_m^2).
\end{equation}
In the limit $\omega_m \ll B$ we get an effective Hamiltonian describing uncoupled (in the first order) effective qubit and mechanics, and including the mechanical Stark shift,
\begin{equation}
H_{\rm eff}=U_{SW} H U_{SW}^\dagger \approx H_{qb} + \omega_m a^\dagger a - \frac{g_m^2}{B} x_m^2 \sigma_3 + o(g_m^3).
\end{equation}
This transformation also changes the qubit operators, in particular for the ladder operators
\begin{equation}
\tilde \sigma_{\uparrow/\downarrow} \equiv U_{\rm SW}\sigma_{\uparrow/\downarrow} U_{\rm SW}^\dagger = \sigma_{\uparrow/\downarrow} - \frac{g_m}{B} x_m \sigma_3 - \frac{g_m^2}{B^2} x_m^2 \sigma_1 + o(g_m^3).
\end{equation}
Now, in order to find out the effect of qubit dissipation on the effective mechanical coordinate, we should use these transformed ladder operators in $\Gamma(t)$. We get eight terms,
\begin{subequations}
\begin{align}
T_1 &= \sum_j \lambda_j^2 \int \frac{d\omega}{2\pi} S_j(\omega)  \int_{t_0}^t dt' e^{-i\omega t'}  \tilde \sigma_\uparrow(t) \tilde \sigma_\downarrow(t_1) \rho_q(t)\\
T_2 &= -\sum_j \lambda_j^2 \int \frac{d\omega}{2\pi} S_j(\omega)  \int_{t_0}^t dt' e^{-i\omega t'}  \tilde \sigma_\uparrow(t_1) \rho_q(t) \tilde \sigma_\downarrow(t) \\
T_3 &= \sum_j \lambda_j^2 \int \frac{d\omega}{2\pi} S_j(\omega)  \int_{t_0}^t dt' e^{-i\omega t'}  \tilde \sigma_\downarrow(t) \tilde \sigma_\uparrow(t_1) \rho_q(t)\\
T_4 &= -\sum_j \lambda_j^2 \int \frac{d\omega}{2\pi} S_j(\omega)  \int_{t_0}^t dt' e^{-i\omega t'}  \tilde \sigma_\downarrow(t_1) \rho_q(t) \tilde \sigma_\uparrow(t) \\
T_5 &= \sum_j \lambda_j^2 \int \frac{d\omega}{2\pi} S_j(-\omega)  \int_{t_0}^t dt' e^{-i\omega t'} \rho_q(t) \tilde \sigma_\uparrow(t_1) \tilde \sigma_\downarrow(t) \\
T_6 &= -\sum_j \lambda_j^2 \int \frac{d\omega}{2\pi} S_j(-\omega)  \int_{t_0}^t dt' e^{-i\omega t'}  \tilde \sigma_\uparrow(t) \rho_q(t) \tilde \sigma_\downarrow(t_1) \\
T_7 &= \sum_j \lambda_j^2 \int \frac{d\omega}{2\pi} S_j(-\omega)  \int_{t_0}^t dt' e^{-i\omega t'} \rho_q(t) \tilde \sigma_\downarrow(t_1) \tilde \sigma_\uparrow(t) \\
T_8 &= -\sum_j \lambda_j^2 \int \frac{d\omega}{2\pi} S_j(-\omega)  \int_{t_0}^t dt' e^{-i\omega t'}  \tilde \sigma_\downarrow(t) \rho_q(t) \tilde \sigma_\uparrow(t_1).
\end{align}
\end{subequations}
Noting that the replacement $B \mapsto -B$ is equivalent to flipping
the spins, we have $T_3(B)=T_1(-B)$, $T_7(B)=T_5(-B)$,
$T_2(B)=T_3(-B)$, and $T_8(B)=T_6(-B)$. Moreover, because $\tilde
\sigma_\downarrow(t)$ is a complex conjugate of $\tilde
\sigma_\uparrow(t)$, the relevant terms contain only the time difference $t-t_1=t'$. Therefore interchanging the time indices amounts to changing the sign of the frequency. Because of this, we have $T_8=T_4$ and $T_6=T_2$. 

It is therefore enough to calculate only $T_1, T_4$ and $T_5$. For this we need to calculate the time dependence of the correction terms in the interaction picture. We disregard the resonator Stark shift since it is small compared to the bare qubit frequency. Then, we have
\begin{equation}
\tilde \sigma_{\uparrow/\downarrow}(t) = \sigma_{\uparrow/\downarrow}(t) - \frac{g_m}{B} (a e^{-i\omega_m t} + a^\dagger e^{i\omega_m t}) \sigma_3 - \frac{g_m^2}{B^2}(2 a^\dagger a +1 + a^2 e^{-2i\omega_m t} + (a^\dagger)^2 e^{2i\omega_m t})(\sigma_\uparrow(t)+\sigma_\downarrow(t)).
\end{equation}
We calculate the dissipator to the second order in $g_m$. The first-order term induces a weak renormalization in the mechanical resonance frequency and we neglect it in the following. The second-order term is a dissipative term. We get $T_j=T_{ja}+T_{jb}$, where
\begin{subequations}
\begin{align}
T_{1a}&=\sum_j \lambda_j^2 \frac{g^2}{B^2} 2S_j(-B) n_q(B) (2 a^\dagger a+1) \rho_m(t)\\ 
T_{1b}&=\sum_j \lambda_j^2 \frac{g^2}{B^2} \left[S_j(-\omega_m) a a^\dagger \rho_m(t) + S(-\omega_m) a^\dagger a \rho_m(t)\right]\\
T_{4a}&=-\sum_j \lambda_j^2 \frac{g^2}{B^2} S_j(-B) n_q(B) \{2 a^\dagger a+1, \rho_m(t)\}\\
T_{4b}&= \sum_j \lambda_j^2 \frac{g^2}{B^2} \left[S_j(-\omega_m) a^\dagger \rho_m(t) a + S(-\omega_m) a \rho_m(t) a^\dagger\right]\\
T_{5a}&=\sum_j \lambda_j^2 \frac{g^2}{B^2} 2S_j(-B) n_q(B) \rho_m(t) (2 a^\dagger a+1)\\ 
T_{5b}&=\sum_j \lambda_j^2 \frac{g^2}{B^2} \left[S_j(-\omega_m) \rho_m(t) a a^\dagger  + S(-\omega_m) \rho_m(t) a^\dagger a\right].
\end{align}
\end{subequations}
Here $n_q(B)\equiv {\rm Tr}_q[\rho_q \sigma_\uparrow \sigma_\downarrow]$ is the probability of occupying the excited state of the qubit ($n_q(-B)$ is the probability of the ground state). We hence note that $T_{1a}+T_{5a}=-2T_{4a}$ and moreover, the $b$-terms are independent of $B$. The total qubit-induced dissipator is
\begin{equation}
\begin{split}
{\cal L}_{qm}&=[T_1(B)+T_1(-B)+T_5(B)+T_5(-B)+2 T_4(B)+2 T_4(-B)]=2[T_{1b}+T_{5b}+2 T_{4b}]\\
&-\underbrace{2 \frac{g^2}{B^2}\sum_j \lambda_j^2 S_j(\omega_m)}_{\gamma_{rel}/2} [2a \rho_m(t) a^\dagger-{\rho,a^\dagger a}] - \underbrace{2 \frac{g^2}{B^2}\sum_j \lambda_j^2 S_j(-\omega_m)}_{\gamma_{exc}/2} [2a^\dagger \rho_m(t) a-{\rho,a a^\dagger }].
\end{split}
\end{equation}
We hence finally found the regular oscillator dissipator. 

Let us first consider the case of thermal noise in the qubit bath. Because an equilibrium correlator satisfies the detailed balance relation $S_j(\omega)=\exp(\omega/T_q) S_j(-\omega)$, so do the relaxation and excitation rates (here $T_q$ is the temperature of the qubit bath). Moreover, we can use the fluctuation-dissipation theorem for each bath oscillator separately,
\begin{equation}
S_j(\omega)=\gamma_j \omega \left[\coth\left(\frac{\omega}{2 T_q}\right)+1\right],
\end{equation}
where $\gamma_j$ is the dissipative part of the response coefficient. Let us assume $\gamma_j$ to be independent of frequency. In that case, we can write
\begin{equation}
\gamma^q_{\rm rel}= 4 \sum_j \lambda_j^2 \gamma_j \frac{B}{1-e^{-B/T_q}} \overset{T_q \ll B} \approx 4 B \sum_j \gamma_j \lambda_j^2.
\end{equation}
We can hence express the effective oscillator relaxation due to hybridization as
\begin{equation}
\gamma_{\rm rel}=2 \gamma_{\rm rel}^q \frac{g_m^2 \omega_m}{B^3} \frac{1}{1-e^{-\omega_m/T}} = 2 \gamma_{\rm rel}^q \frac{g_m^2 \omega_m}{B^3} (1+n_m),
\end{equation}
where $n_m=(\exp(\omega_m/T)-1)^{-1}$ is the Bose function at frequency $\omega_m$.

For $1/f$ (flicker) noise, the noise in the bath increases with a decreasing frequency. In that case the dissipation rate satisfies
\begin{equation}
\frac{\gamma_{\rm rel}}{\gamma_{\rm rel}^q} = \frac{2g_m^2}{B^2}\frac{\sum_j \lambda_j S_j(\omega_m)}{\sum_j \lambda_j S_j(B)} \approx \frac{2g_m^2}{B \omega_m},
\end{equation}
which is not necessarily small any more.

\section{Cavity dephasing}
Let us then study the $x^2\sigma_1$ coupling to the cavity (coupling
terms $g_{q1}$ and $g_{q2}$ in Eqs.~(6-7) of the main text),
\begin{equation}
H = \frac{B}{2}\sigma_3 +\sum_j\bigg[ \omega_j b^{\dag}b +\lambda_j(b^{\dag}+b)\sigma_1\bigg] + \omega_c a^{\dag}a + g\big((a^{\dag})^2+a^2 +2a^{\dag}a+1\big)\sigma_1,
\end{equation}
We first decouple the cavity from the qubit with the Schrieffer-Wolff -transformation $U=e^S$ where
\begin{equation}
S=-\frac{2g\omega_c}{B^2-(2\omega_c)^2}\big((a^{\dag})^2-a^2\big)\sigma_1 + \frac{iBg}{B^2-(2\omega_c)^2}\big((a^{\dag})^2+a^2\big)\sigma_2 + \frac{ig}{B}(2a^{\dag}a+1)\sigma_2.
\end{equation}
We write $\sigma_{\pm}$ in the first order in $g_3^m/\omega_m$. We assume that the second order does not affect the dissipator, similar to the $x\sigma_1$-coupling. 
\begin{equation}
\sigma_{\pm}(t) \rightarrow \frac{g}{B}\big((a^{\dag})^2e^{2i\omega_ct}+a^2e^{-2i\omega_ct}\big)\sigma_{3} +\frac{g}{B}(2a^{\dag}a+1)\sigma_3.
\end{equation}
With this representation, one can repeat the calculation as done above
for the $x\sigma_1$ coupling. Again, it is enough to calculate
\begin{equation}
\begin{split}
T_1=&\sum_j \lambda_j^2\bigg[\frac{g^2}{B^2}\bigg((a^{\dag})^2a^2S_j(2\omega_m) +a^2(a^{\dag})^2S_j(-2\omega_m)\bigg)+\frac{g^2}{B^2}(2a^{\dag}a+1)^2S_j(0)\bigg]\rho_c(t)\\
T_4=&-\sum_j \lambda_j^2\bigg[\frac{g^2}{B^2}\bigg((a^{\dag})^2\rho_c(t)a^2S_j(-2\omega_m) +a^2\rho_c(t)(a^{\dag})^2S_j(2\omega_m)\bigg)\\
&+\frac{g^2}{B^2}(2a^{\dag}a+1)\rho_c(t)(2a^{\dag}a+1)S_j(0)\bigg]\\
T_5=&\sum_j \lambda_j^2\rho_c(t)\bigg[\frac{g^2}{B^2}\bigg((a^{\dag})^2a^2S_j(2\omega_c) +a^2(a^{\dag})^2S_j(-2\omega_c)\bigg)+\frac{g^2}{B^2}(2a^{\dag}a+1)^2S_j(0)\bigg],
\end{split}
\end{equation}
where we have traced over the qubit degree of freedom and neglected
terms that are linear in $B$, since $T_3(B)=T_1(-B)$, etc. Notice that
the above equations correspond to $T_{1b},T_{4b}$ and $T_{5b}$ in the
previous section. By employing the relations for $T$'s obtained there, we obtain the dissipator
\begin{equation}
\begin{split}
\mathcal{L} =& -[T_1(B)+T_1(-B)+T_5(B)+T_5(-B)+2T_4(B)+2T_4(-B)]\\
=& \overbrace{2\frac{g^2}{B^2}\sum_j\lambda_j^2 S_j(2\omega_c)}^{\gamma_{\rm rel}/2}\bigg(2a^2\rho_c (a^{\dag})^2-\rho_c (a^{\dag})^2a^2-(a^{\dag})^2a^2\rho_c\bigg)\\
&+ \overbrace{2\frac{g^2}{B^2}\sum_j\lambda_j^2 S_j(-2\omega_c)}^{\gamma_{\rm exc}/2}\bigg(2(a^{\dag})^2\rho_c a^2-\rho_c a^2(a^{\dag})^2-a^2(a^{\dag})^2\rho_c\bigg)\\
&+ \overbrace{2\frac{g^2}{B^2}\sum_j\lambda_j^2 S_j(0)}^{\gamma_{\rm \phi}/4}\bigg(2(a^{\dag}a+aa^{\dag})\rho_c (a^{\dag}a+aa^{\dag})-\rho_c (a^{\dag}a+aa^{\dag})^2-(a^{\dag}a+aa^{\dag})^2\rho_c\bigg).
\end{split}
\end{equation}
Above, the two first lines describe two-photon relaxation/excitation
processes of the cavity via the coupling to the qubit bath. They have
otherwise similar rates as that for the linear coupling, but instead
of $S(\pm \omega_c)$ they are proportional to $S(\pm 2\omega_c)$. The
last row is a dephasing term for the cavity, which, for example, leads into an
increase of the linewidth of $\langle a^{\dag}\rangle$ by
$\gamma_{\phi}$. 

In qubits the pure dephasing rate is \cite{heikkilabook}
$\gamma_{\phi}^q=2\sum_j \lambda_j^2 S_j(0)$, except that the coupling
in that case is to $\sigma_3$, and not to $\sigma_1$ as in
Eq.~\eqref{eq:ham} above. Assuming that the coupling to the bath is
isotropic (of the same order for the coupling perpendicular and
parallel with the qubit Hamiltonian), the additional cavity
dephasing due to the hybridization is hence $\gamma_{\phi}\sim g^2
\gamma_\phi^q/B^2$. 

\section{Cross-Kerr coefficient}

In the main text we describe the cross-Kerr coefficient obtained from the Josephson inductance approach in the case of a symmetric system ($E_{J1}=E_{J2}$) and in the absence of a flux ($\phi_a=0$). Lifting these assumptions, we get from the expansion of the effective qubit field
\begin{equation}
\hbar g_{cK}=\frac{8 g_m^2 \left(B^2 \left(B^2-3 B_3^2\right) \left(B_1 g_{q1}+B_2
   g_{q2}+g_{c1}^2+g_{c2}^2\right)-3 \left(B^2-5 B_3^2\right) \left(B_1 g_{c1}+B_2
   g_{c2}\right){}^2\right)}{B^7}.
\end{equation}
This applies in the weak coupling limit where the Stark shifts of the
cavity and the mechanics are negligible. For a strong cavity Stark
shift, i.e., when the Josephson inductance is of the order of $L$ or
smaller, the expression should be renormalized. This is easiest to do by directly using the Josephson inductance approach and expanding the effective cavity frequency into the second order in the mechanical coordinate. In other words, we are expanding the radiation pressure coupling to the second order. This yields three terms, (the factor four comes from the rotating wave approximations relating $x_c^2 x_m^2 \approx 4 c^\dagger c a^\dagger a$).
\begin{equation*}
g_{cK}=4x_{ZP}^2 \partial_x^2 \omega_c = 2x_{ZP} \partial_x g_{rp}=4\left[(\partial_x \partial_{L_J^{-1}} \omega_c ) \partial_{n_g} L_{J}^{-1} \partial_x n_g +\partial_{L_J^{-1}} \omega_c (\partial_x \partial_{n_g} L_{J}^{-1}) \partial_x n_g + \partial_{L_J^{-1}} \omega_c  \partial_{n_g} L_{J}^{-1} \partial_x^2 n_g\right].
\end{equation*}
The last term is proportional to $V_g \partial_x^2 C_g$, and hence much smaller than the other terms that are proportional to $V_g^2$. The two other terms are
\begin{equation}
\label{eq:gck}
g_{cK}=\omega_c \left(\frac{V_g \partial_x C_g}{2e}\right)^2 \frac{L}{LL_J^{-1}+1}\left[\frac{3L}{LL_J^{-1}+1} (\partial_{n_g}L_J^{-1})^2-2\partial_{n_g}^2 L_J^{-1}\right].
\end{equation}
It is the second term inside the square brackets that describes the weak-coupling limit obtained from the Schrieffer-Wolff approach. It indeed dominates the first term for $L_J \gg L$. The overall prefactor $1/(LL_J^{-1}+1)=(\omega_c/\omega_{c}^0)^2$ is a renormalization effect that would be obtained by applying the Bogoliubov transformation on the cavity operators in the cross-Kerr term. However, due to the first term there are also other renormalization effects. In practice, these renormalizations yield the terms proportional to $l$ in Eq.~(9) of the main text. 

In the case of asymmetric junctions and finite flux, the resulting expression for $g_{cK}$ is too long to be written here. However, it can be straigthforwardly obtained from combining Eq.~\eqref{eq:gck} and $L_J^{-1}=(2e)^2/\hbar^2 [\partial_\phi^2 E_{\rm SCPT}(\phi)]$ with $E_{\rm SCPT}(\phi)$ given below Eq.~(4) in the main text.

 
\section{Spectrum of the full Hamiltonian}
The effective cavity-mechanics Hamiltonian, where the systems are
coupled via the qubit is of the form 
\begin{equation}
H_{\rm eff}=\hbar \omega_c d^\dagger d + \hbar \omega_m b^\dagger b +
\hbar g_0 d^\dagger d (b^\dagger+b)+g_{cK} d^\dagger d b^\dagger b,
\end{equation}
where we included the cross-Kerr coupling in the Hamiltonian (10) of
the main text. 
The spectrum of this Hamiltonian is described by two quantum
numbers. The first one is the occupation number $n_d$ of the (effective)
cavity, because $H_{\rm eff}$ commutes with $\hat n_d = d^\dagger
d$. The Hamiltonian thus separates into different blocks specified by
$n_d=0, 1, 2, \dots$. The Hamiltonian for block $n_d$ is thus
\begin{equation}
H_{n_d}=n_d \hbar \omega_c +\hbar (\omega_m+n_d g_{cK})b^\dagger b +
\hbar g_0 n_d
(b^\dagger+b).
\end{equation}
This can be diagonalized by introducing a translation
\begin{equation}
b=e-\frac{g_0 n_d}{\omega_m+n_d g_{cK}},
\end{equation}
where $e$ is another bosonic annihilation operator. The resulting
Hamiltonian $H_{n_d}$ is then diagonal, and commutes with $e^\dagger
e$. We hence get another good quantum number $n_e$ labeling the
eigenstates within $H_{n_d}$. The eigenstates are thus
$|n_d,n_e\rangle$ with energies
\begin{equation}
\epsilon_{n_d,n_e} = n_d \hbar \omega_c + n_e \hbar (\omega_m+n_d
g_{cK})-\hbar \frac{g_0^2 n_d^2}{\omega_m+n_d g_{cK}}.
\end{equation}
This spectrum is schematically depicted in
Fig.~\ref{fig:spectrum}. Note that for each $n_d$, we describe 
linear harmonic oscillator states (corresponding mostly to the mechanical
vibrations), but the spacing of these states depends on
$n_d$ due to the cross-Kerr coupling $g_{cK}$. On the other hand, for
a given $n_e$, the effective cavity is an anharmonic oscillator with a
harmonic term having the level spacing $\hbar(\omega_c + n_e g_{cK})$ and an
anharmonic term of the form
\begin{equation}
-\hbar \frac{g_0^2 \hat n_d^2}{\omega_m + n_d g_{cK}} \approx -\hbar \frac{g_0^2
  \hat n_d^2}{\omega_m}.
\end{equation}
This spectrum is a good starting point for the analysis of the physics
of a driven system.

\begin{figure}[h]
\centering
\includegraphics[width=\linewidth]{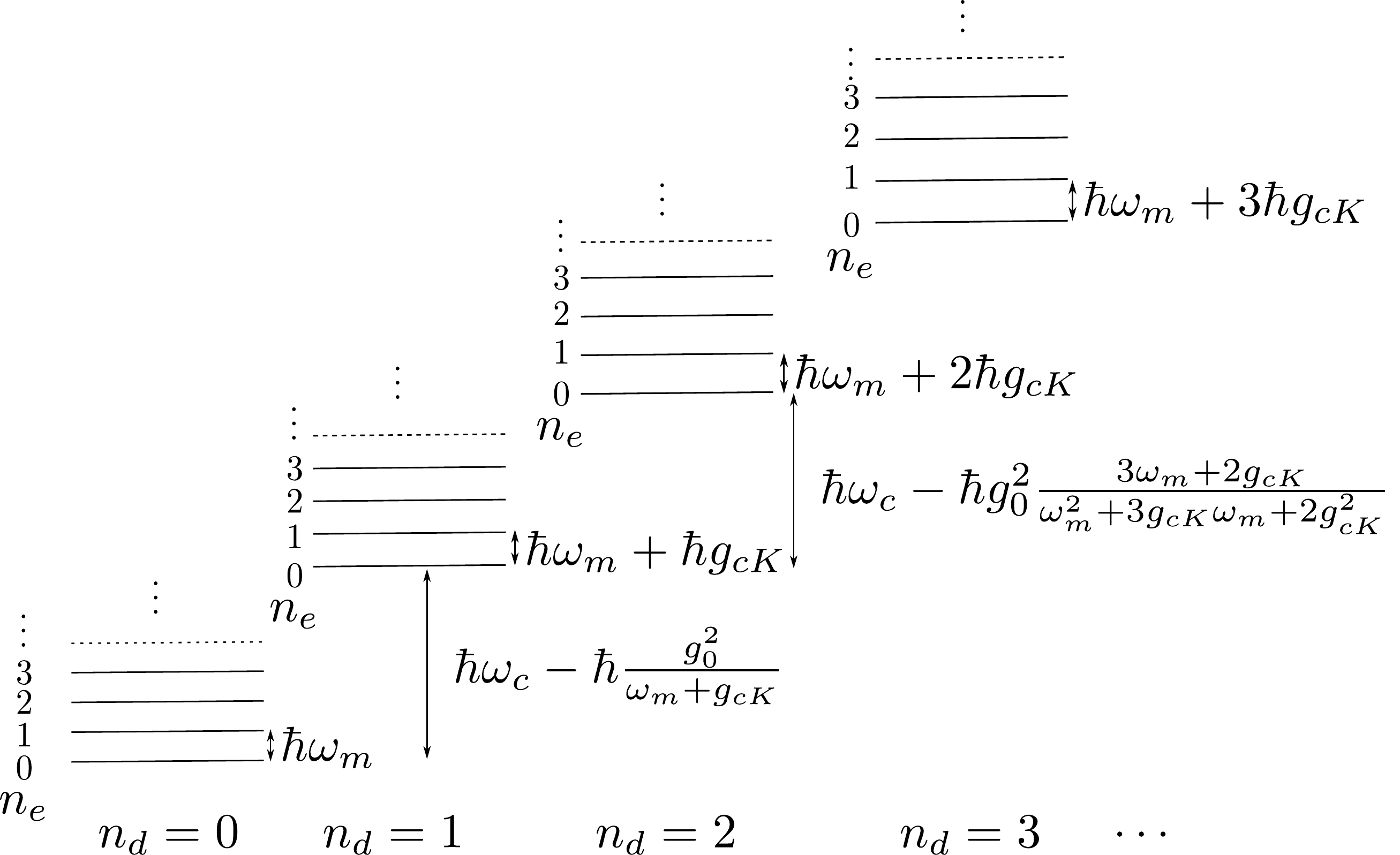}
\caption{Schematic spectrum of the effective Hamiltonian, Eq.~(11) of
  our main text.}
\label{fig:spectrum}
\end{figure}

\end{widetext}

\begin{thebibliography}{99}
\bibitem{teufel11} J.D. Teufel, T. Donner, D. Li, J.W. Harlow,
  M.S. Allman, K. Cicak, A.J. Sirois, J.D. Whittaker, K.W. Lehnert,
  and R.W. Simmonds, Nature {\bf 475}, 359 (2011).
\bibitem{painter11} J. Chan, T.P.M. Alegre, A.H. Safavi-Naeini,
  J.T. Hill, A. Krause, S. Gr\"oblacher, M. Aspelmeyer, and O. Painter,
  Nature {\bf 478}, 89 (2011).
\bibitem{safavinaeni12} A.H. Safavi-Naeini, J. Chan, J.T. Hill,
  Thiago P Mayer Alegre, A. Krause, and O. Painter, Phys. Rev. Lett. {\bf
    108}, 033602 (2012).
\bibitem{massel11} F. Massel, T.T. Heikkil\"a, J.M. Pirkkalainen,
  S.U. Cho, H. Saloniemi, P.J. Hakonen, and M.A. Sillanp\"a\"a, Nature
  {\bf 480}, 351 (2011).
\bibitem{mancini97} S. Mancini, V.I. Man'ko, and P. Tombesi,
  Phys. Rev. A {\bf 55}, 3042 (1997).
\bibitem{bose97} S. Bose, K. Jacobs, and P.L. Knight, Phys. Rev. A {\bf
    56}, 4175 (1997).
\bibitem{ludwig08} M. Ludwig, B. Kubala, and F. Marquardt, New
  J. Phys. {\bf 10}, 095013 (2008).
\bibitem{nunnenkamp11} A. Nunnenkamp, K. B\o{}rkje, and S.M. Girvin,
  Phys. Rev. Lett. {\bf 107}, 063602 (2011).
\bibitem{rabl11} P. Rabl, Phys. Rev. Lett. {\bf 107}, 063601 (2011).
\bibitem{qian12}  J. Qian, A.A. Clerk, K. Hammerer, and F. Marquardt,
  Phys. Rev. Lett. {\bf 109}, 253601 (2012).
\bibitem{xu12} X-W. Xu, H. Wang, J. Zhang, and Y.X. Liu,
  arXiv:1210.0070.
\bibitem{nunnenkamp12} A. Nunnenkamp, K. B\o{}rkje, and S.M. Girvin,
  Phys. Rev. A {\bf 85}, 051803 (2012).
\bibitem{stannigel12} K. Stannigel, P. Komar, S.J.M. Habraken,
  S.D. Bennett, M.D. Lukin, P. Zoller, and P. Rabl,
  Phys. Rev. Lett. {\bf 109}, 013603 (2012).
\bibitem{xu13} G.-F. Xu and C.K. Law, Phys. Rev. A {\bf 87}, 053849
  (2013).
\bibitem{kronwald13} A. Kronwald and F. Marquardt,
  Phys. Rev. Lett. {\bf 111}, 133601 (2013).
\bibitem{akram13} U. Akram, W.P. Bowen, and G.J. Milburn, New
  J. Phys. {\bf 15}, 093007 (2013).
\bibitem{liao13} J.-Q. Liao and F. Nori, Phys. Rev. A {\bf 88}, 023853 (2013).
\bibitem{murch08} K.W. Murch, K.L. Moore, S. Gupta, and
  D.M. Stamper-Kurn, Nat Phys {\bf 4}, 561 (2008).
\bibitem{brennecke08} F. Brennecke, S. Ritter, T. Donner, and
  T. Esslinger, Science {\bf 322}, 235 (2008).
\bibitem{chan12} J. Chan, A.H. Safavi-Naeini, J.T. Hill, S. Meenehan,
  and O. Painter, Appl. Phys. Lett. {\bf 101}, 081115 (2012).
\bibitem{xuereb12} A. Xuereb, C. Genes, and A. Dantan,
  Phys. Rev. Lett. {\bf 109}, 223601 (2012).
  \bibitem{Nori06}  C. P. Sun, L. F. Wei, Y.X. Liu, and F. Nori, Phys. Rev. A, {\bf 73}, 022318 (2006).
  \bibitem{buks2007}  E. Buks, S. Zaitsev, E. Segev, B. Abdo, and M. P. Blencowe, Phys. Rev. E {\bf 76}, 026217 (2007).
\bibitem{nori2009}  J. Zhang,Y. X. Liu, and F. Nori, Phys. Rev. A {\bf 79}, 052102 (2009).
\bibitem{pirkkalainen12} J.M. Pirkkalainen, S.U. Cho, J. Li,
  G.S. Paraoanu, P.J. Hakonen, and M.A. Sillanp\"a\"a, Nature {\bf 494}, 211 (2013).
\bibitem{sillanpaa04} M.A. Sillanp\"a\"a, L. Roschier, and P.J. Hakonen,
  Phys. Rev. Lett. {\bf 93}, 066805 (2004).
\bibitem{nakamura} Y. Nakamura, Yu.A. Pashkin, and J.S. Tsai, Nature {\bf 398},
786 (1999).
\bibitem{quantronium} D. Vion \emph{et al.}, Science {\bf 296}, 886 (2002).
\bibitem{makhlin01} Yu. Makhlin, G. Sch\"on, and A. Shnirman,
    Rev. Mod. Phys. {\bf 73}, 357 (2001).
\bibitem{dimensionnote} We use the convention that the quantities
  related to qubits ($B_j$ fields and couplings) are in units of
  energy, whereas those related only to oscillators are in units of
  frequency. 
\bibitem{regal08} C.A. Regal, J.D. Teufel, and K.W. Lehnert, Nat. Phys.
  {\bf 4}, 555 (2008).
  \bibitem{lahaye09} M.D. LaHaye, J. Suh, P.M. Echternach, K.C. Schwab,
  and M.L. Roukes, Nature {\bf 459}, 960 (2009).
\bibitem{minimumnote} Typically there is a unique minimum, but for a
  complicated form of $E_{\m{SCPT}}(\phi)$ there may be more than one of
  them. In that case we choose the global minimum. The presence of
  multiple minima would show up as hystereris.
\bibitem{sulkko10} J. Sulkko, M.A. Sillanp\"a\"a, P. H\"äkkinen, L. Lechner,
  M. Helle, A. Fefferman, J. Parpia, and P.J. Hakonen, Nano Lett. {\bf
    10}, 4884 (2010).
\bibitem{tian} L. Tian, Phys. Rev. B {\bf 79}, 193407 (2009).
\bibitem{bravyi11} S. Bravyi, D. DiVincenzo, and D. Loss,
  Ann. Phys. {\bf 326}, 2793 (2011).
\bibitem{gramich13} V. Gramich, B. Kubala, S. Rohrer, and
  J. Ankerhold, arXiv:1307.2495.
\bibitem{gardinerzoller} C.W. Gardiner and P. Zoller, {\em Quantum
    Noise}, Springer (Berlin Heidelberg, 2010).
\bibitem{schuster05} D.I. Schuster, A. Wallraff, A. Blais, L. Frunzio,
  R.S. Huang, J. Majer, S.M. Girvin, and R.J. Schoelkopf,
  Phys. Rev. Lett. {\bf 94}, 123602 (2005).
\bibitem{bogoliubovnote} Note that the corrections
  $(\omega_c^0/\omega_c)$ and $(\omega_m^0/\omega_m)^{1/2}$ coming from the Bogoliubov transformation
  are formally of a higher order than assumed in the coupling
  expansion. The first term is necessary to properly account for the effect
  of the Josephson inductance in the limit $L_J \lesssim L$.
\bibitem{gunnarsson08} D. Gunnarsson, J. Tuorila, A. Paila,
  J. Sarkar, E. Thuneberg, Yu. Makhlin, and P. Hakonen,
  Phys. Rev. Lett. {\bf 101}, 256806 (2008).
\end{thebibliography}

\begin{thebibliography}{9}
\bibitem{heikkilabook} T.T. Heikkil\"a, {\em The Physics of
    Nanoelectronics} (Oxford University Press 2013).
\end{thebibliography}
\end{document}